\documentclass[runningheads]{llncs}
\usepackage[utf8]{inputenc}
\usepackage[english]{babel}
\usepackage[sort&compress, square, numbers]{natbib}
\usepackage{pxfonts}
\usepackage{graphicx}
\usepackage{hyperref}

\title{Chapter 41: Identifying stimulus-driven neural activity patterns in multi-patient intracranial recordings}
\author{Jeremy R. Manning}
\titlerunning{Identifying stimulus-driven neural activity}
\authorrunning{J. R. Manning}

\institute{Dartmouth College, Hanover, NH, USA\\\email{jeremy.r.manning@dartmouth.edu}}

\begin{document}
\maketitle
\begin{center}
Note: This chapter is forthcoming in \textit{Intracranial EEG for Cognitive Neuroscience}\\
Draft current as of \today
\end{center}

\begin{abstract}
  Identifying stimulus-driven neural activity patterns is critical for
  studying the neural basis of cognition. This can be particularly
  challenging in intracranial datasets, where electrode locations
  typically vary across patients. This chapter first presents an
  overview of the major challenges to identifying stimulus-driven
  neural activity patterns in the general case.  Next, we will review
  several modality-specific considerations and approaches, along with
  a discussion of several issues that are particular to intracranial
  recordings.  Against this backdrop, we will consider a variety of
  within-subject and across-subject approaches to identifying and
  modeling stimulus-driven neural activity patterns in multi-patient
  intracranial recordings.  These approaches include generalized
  linear models, multivariate pattern analysis, representational similarity analysis, joint
  stimulus-activity models, hierarchical matrix factorization models,
  Gaussian process models, geometric alignment models, inter-subject
  correlations, and inter-subject functional correlations.  Examples
  from the recent literature serve to illustrate the major concepts
  and provide the conceptual intuitions for each approach.

\end{abstract}

\renewcommand{\thesection}{41.\arabic{section}}

\section{Overview}
Studying brain function often requires identifying brain responses to
a given stimulus or set of stimuli.  For some stimuli, and for some
systems, this identification problem is relatively straightforward.
For example, when a photopigment in a retinal photoreceptor absorbs
light, this triggers a cascade of responses that is ultimately sent
from the retina to other brain areas via the optic
nerve~\citep{Jaco21}.  In the general sense, however (i.e., for
arbitrarily complex stimuli and arbitrary brain areas), the problem of
identifying neural responses to known (or unknown) stimuli can be
incredibly challenging~\citep{JonaKord17}.

\subsection{Why is it challenging to identify stimulus-driven brain
  activity?}
To illustrate the enormity of the challenge of identifying
stimulus-driven brain responses in the general sense, it can be useful
to start by considering what form a complete solution might take.
First, we need some means of defining (and measuring) what brain
activity \textit{is}.  For example, should we concern ourselves with
measuring membrane potentials or firing rates of individual neurons?
Or population activity in a given brain structure or network?  And is
it more appropriate to analyze or interpret activity patterns in the
\textbf{time domain} (e.g., firing rate or voltage as a function of
time), or in the \textbf{frequency domain} (e.g., characterizing the
signal through the relative contributions of its constituent
sinusoidal components at different frequencies)?  Should we consider
neurons and/or brain structures in isolation, or should we instead
interpret each ``unit'' of activity within the context of the
network(s) it participates in or contributes to?  We discuss several
different approaches to these questions (and their relative
trade-offs) in Section~\ref{sec:activity}.

Second, we need some means of characterizing (and ideally,
quantifying) the stimulus itself.  For a simple stimulus, such as a
single photon of light, emitted from a known location in an otherwise
completely dark room, constructing a sufficiently comprehensive model
of the stimulus might be straightforward-- and perhaps even trivial.
For other stimuli, such as real-world experiences, constructing a
comprehensive model of ``what is happening'' can be highly complex (at
best).  Essentially, building a stimulus model entails quantifying how
different \textbf{features}, or properties, change over time.  In our
single-photon example, we might represent the stimulus as a timeseries
of zeros (no photon present) and ones (photon present).  For more
complex stimuli, however, it may not even be clear what the features
\textit{are}.  We discuss considerations and approaches to building
explicit stimulus models in Section~\ref{sec:stimmodel}.

Characterizing brain activity and building stimulus models are each
complex challenges in their own right.  Linking the two provides its
own set of additional challenges.  We discuss these issues in
Section~\ref{sec:linking}.

\subsection{How can we measure neural ``activity'' in the human
  brain? \label{sec:activity}}

The brain is a complex organ comprising myriad cell types that
interconnect to form a vast network.  When neuroscientists use the term
\textit{brain activity}, this can refer to a variety of possible
physical and physiological phenomena.  To contextualize what brain
activity means, it can be helpful to first consider the brain's structure
and function.

\textbf{Neurons} are the dominant cell type in the brain; the adult
human brain contains roughly 100 billion neurons.  The
\textbf{cerebral cortex}, commonly associated with high-level brain
function, comprises roughly 80\% of the adult human brain's mass, but
only roughly 20\% of its neurons~\citep{Herc09}.  In general,
\textbf{brain activity} refers to changes in cellular processes that
neurons undergo.  These processes can take many forms,
including~\citep{KandEtal00}: rapid changes in membrane voltage that
result in neurotransmitter release, called \textbf{action
  potentials}~\citep{HodgHuxl52}; sub-threshold changes in the
membrane voltages of individual neurons or populations of neurons;
neurotransmitter release; changes in the composition or distribution
of cell surface proteins (including ion channels); metabolic changes
(such as increased or decreased blood flow); changes in ion
concentrations; structural or anatomical changes, and so on.  Because
the term brain activity often refers to the activities of neurons
specifically, the term \textbf{neural activity} can be a more precise
way of referring to these phenomena.  However, it is worth noting that
neurons are not the \textit{only} cell type in the brain.  For
example, the brain also contains roughly 100 billion \textbf{glial
  cells}~\citep{AzevEtal09, MotaHerc14} , which also play an important
role in supporting the neuronal function, synapse formation, and brain
metabolism~\cite{Barr08}.

\begin{figure}[tp]
\centering
\includegraphics[width=\textwidth]{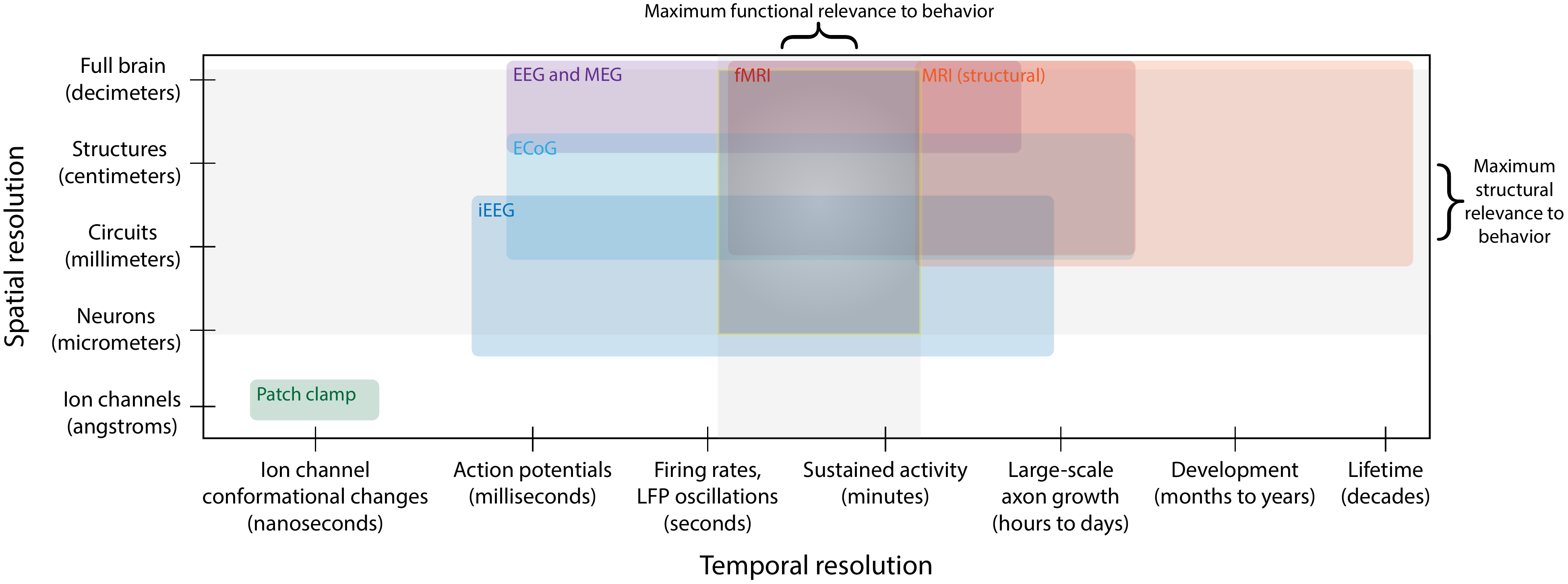}
\caption{\textbf{Spatial versus temporal resolution.}  Each colored
  region represents the temporal ($x$-axis) and spatial ($y$-axis)
  limits of a recording method or neuroimaging modality~\citep[figure
  inspired by][]{SejnEtal14, ChurSejn88}.  Green shading denotes
  \textit{in vitro} methods.  Blue shading denotes invasive \textit{in
    vivo} methods.  Purple, red, and orange denote non-invasive
  \textit{in vivo} methods.  The gray shading denotes suggested ranges of
  maximum structural and functional relevance to experimental-scale
  behaviors.  The shaded region outlined in yellow denotes overlap
  between the horizontal and vertical gray regions.  Note: axes are
  not drawn to scale.}
\label{fig:spacetime}
\end{figure}

After considering the various forms that brain (neural) activity can
take, it can also be useful to define a relevant spatiotemporal scale
(Fig.~\ref{fig:spacetime}).  For example, a biologist concerned with
the structure and function of individual ion channels embedded in the
neuronal cell membrane may be most interested in processes that happen
over a span of picoseconds or nanoseconds (e.g., the amount of time it
takes for an ion to pass through an ion channel, or the amount of time
it takes for an ion channel to change its conformation).  They may
also be most interested in spatial resolutions on the orders of
angstroms (e.g., the approximate scale of an individual ion channel).
At another extreme, a neuroanatomist studying the comparative anatomy
of different species of primates might be most interested in
timescales on the order of decades (e.g., an entire lifetime) and
spatial scales on the order of decimeters (e.g., the size of an adult
brain).  As summarized in Figure~\ref{fig:spacetime}, different
neuroimaging approaches are each associated with a range of temporal
and spatial scales that they are best suited to measure.

If we are specifically interested in stimulus-driven neural activity,
this implies focusing in on a limited range of spatiotemporal
resolutions (gray shading in Fig.~\ref{fig:spacetime}).  Neuroimaging
approaches that enable insights at those resolutions may provide
particularly useful measures of stimulus-driven neural activity.

\subsubsection{Non-invasive approaches.}
Neuroimaging approaches that rely on measurements taken using sensors
that are placed without requiring surgery are referred to as
\textbf{non-invasive}.  In general, non-invasive neuroimaging entails
placing one or more sensors on or near the subject's head.  Examples
include \textbf{scalp electroencephalography} (EEG; i.e., recording
voltages from small electrodes placed on the scalp);
\textbf{magnetoencephalography} (MEG; i.e., measuring tiny changes in
the magnetic field outside of the head caused by neuronal firing); and
\textbf{functional magnetic resonance imaging} (fMRI; i.e., inferring
changes in blood flow associated with neural activity using a powerful
magnet placed around the head).  A related approach, \textbf{magnetic
  resonance imaging} (MRI) uses strong magnetic fields, magnetic field
gradients, and radio waves to produce a static anatomical image of the
brain.  Each of these approaches is widely used by neuroscientists
interested in studying the neural basis of cognition and behavior.  A
benefit of relying on non-invasive neuroimaging is that these
approaches are low-risk and may be safely used on healthy
(non-patient) participants, and without the supervision of a
physician.  The main drawback of non-invasive neuroimaging is that,
because these approaches all rely on sensors placed outside of the
head, any relevant activity that is filtered out by the skull, or that
is too weak to be measured from distant sensors, cannot be captured.
This means that non-invasive neuroimaging approaches tend to have
lower spatiotemporal resolution than invasive approaches.

\subsubsection{Invasive approaches.}
Neuroimaging approaches that require surgery are referred to as
\textbf{invasive}.  Invasive \textit{in vivo} techniques entail
placing sensors directly on the surface of and/or in direct contact
with deep structure inside of a living person's brain.  Examples
include \textbf{intracranial electroencephalography} (iEEG; i.e.,
recording voltages from tiny wires implanted in the brain) and
\textbf{electrocorticography} (ECoG; i.e., recording voltages from
small electrodes lying directly on the brain's cortical surface).
Both iEEG and ECoG are similar to (non-invasive) EEG, in that all
three approaches entail recording aggregate voltages from populations
of many neurons.  The key differences between these approaches are the
locations and sizes of the electrodes.  When sensors are larger and
are placed far from the signal sources (i.e., neurons), as in EEG, the
sensors pick up on relatively large populations of neurons that are
spread over a large portion of the brain.  When sensors are smaller
and placed in direct contact with signal sources, as in ECoG and iEEG,
the sensors pick up on smaller populations of neurons that are closer
to the recording surface of the electrodes.  When tiny
\textbf{microwires} are used to generate iEEG recordings, it is even
possible to record action potentials from individual neurons.

Invasive \textit{in vitro} neuroimaging approaches entails taking
measurements from brain slices or other structures that have been
excised from an intact brain and placed in an isolated environment
such as a petri dish.  For example, \textbf{patch clamp} recordings
use tiny glass micropipettes placed directly on the cell membrane to
capture changes in membrane potential associated with the opening and
closing of individual ion channels.  \textit{In vitro} approaches are
not generally used to study stimulus-driven neural activity in humans,
since excising the to-be-recorded tissues typically isolates the
corresponding neurons from their sensory inputs.  \textit{In vitro}
approaches also require destroying the to-be-recorded tissues, which
presents ethical and safety concerns.

Because invasive neuroimaging approaches require physically cutting
into the brain, they are not appropriate for use in non-patient
populations.  Rather, invasive recordings in humans are typically
taken from neurosurgical patients who have their electrodes implanted
as part of a treatment protocol.  For example, people suffering from
drug-resistant epilepsy may elect to undergo invasive monitoring (from
implanted electrodes) in order to help neurologists localize the most
likely source(s) of their seizures.  During an extended hospital stay,
the patients may elect to participate in research studies that are not
directly related to their treatment, in the interest of advancing
scientific knowledge by providing access to high quality recordings
from their brain.  Stimulus-driven responses in individual neurons or
small circuits that unfold over sub-millisecond timescales can only
be measured using invasive approaches like iEEG and ECoG.  By
providing measurements at both high spatial resolution and high
temporal resolution, intracranial recordings can be ideally suited to
studying stimulus-driven neural activity patterns
(Fig.~\ref{fig:spacetime}).

\subsubsection{Single-channel neural signals.}
\begin{figure}[tp]
\centering
\includegraphics[width=0.8\textwidth]{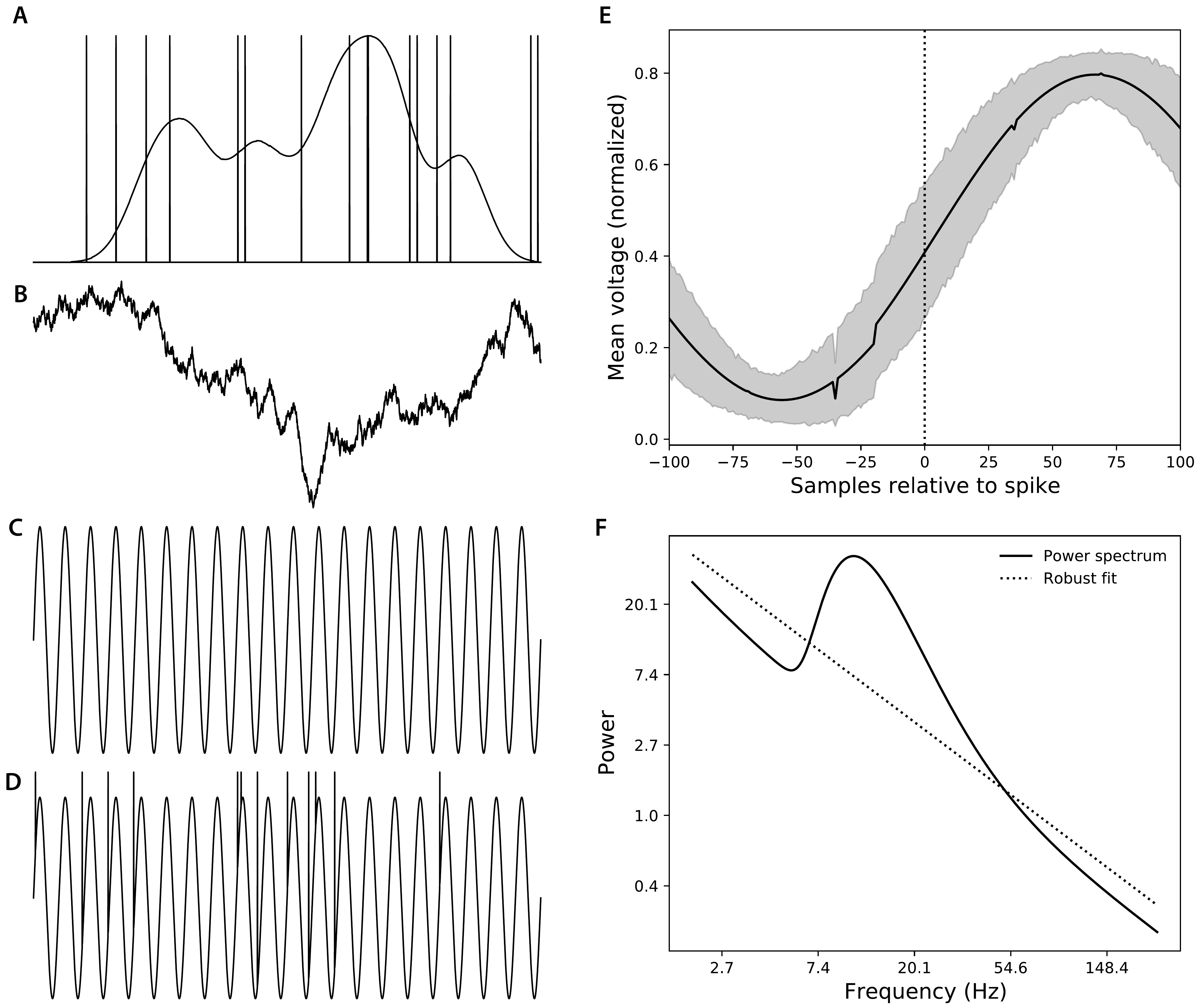}
\caption{\textbf{Measuring and processing single-channel neural
    signal.}  All of the examples shown in this figure are constructed
  using simulated data.  \textbf{A--D.  Neural signals.}  The
  illustrated examples show voltage ($y$-axis, arbitrary units) as a
  function of time ($x$-axis, arbitrary units).  \textbf{A. Neuronal
    firing.}  The vertical lines illustrate the times at which an
  artificial neuron fired action potentials (spikes).  The smooth
  curve shows the timeseries of firing rates, computed by convolving
  the spike timeseries with a Gaussian kernel.  \textbf{B. Local field
    potentials (LFPs).}  LFPs reflect the aggregate neuronal firing
  and sub-threshold changes in membrane potential across thousands of
  neurons near the recording surface of an electrode.
  \textbf{C. Oscillations.}  When the local field potential exhibits
  sinusoidal fluctuations, this can reflect coordinated changes in
  membrane potential across a population of neurons.  \textbf{D. Phase
    coding.}  The timing of an individual neuron's action potentials
  with respect to oscillations in the local field potential can code
  information via the \textit{phase} (angle) relative to a sine wave
  at the oscillation's frequency.  \textbf{E. Spike-triggered
    average.}  Phase coding may be identified by sampling the LFP
  before and after each spike and then averaging across all spikes.
  The spike-triggered average in this panel is computed using the
  artificial LFP displayed in Panel D.  \textbf{F. Power spectrum and
    broadband power.}  Oscillatory contributions to the local field
  potential may be summarized as a \textit{power spectrum} that shows
  the extent to which oscillations at each frequency contribute to the
  LFP.  The power spectrum in this panel is computed using the
  artificial LFP displayed in Panel C.  The underlying height of the
  power spectrum, called \textit{broadband power} may be estimated by
  using robust regression to fit a line to the power spectrum in
  log-log space.  The area under the robust fit line may be used to
  estimate the firing rates of neurons in the underlying
  population~\citep{MannEtal09a}.}
\label{fig:signals}
\end{figure}

A single ECoG or iEEG electrode implanted in a patient's brain
measures changes in membrane potential (voltage) in individual
neurons, other cells and signal sources, and populations of cells.
Because neurons can most effectively transmit signals to other cells
via action potentials, the timings of individual action potentials from a
given neuron, or the firing rate of a neuron, can provide putative
insights into that cell's function (Fig.~\ref{fig:signals}A).  For
example, if a neuron changes its firing rate when the patient is
exposed to a particular stimulus or experience, this could suggest
that the neuron plays some role in processing information pertaining
to that stimulus or experience.

\textbf{Local field potentials} (LFPs) reflect the aggregate neuronal
firing and sub-threshold changes in membrane potential across
thousands of neurons near the recording surface of an electrode
(Fig.~\ref{fig:signals}B).  When LFPs change during exposure to a
stimulus or experience, this can suggest that the underlying
\textit{population} of neurons plays some role in processing that
stimulus or experience.  These changes may be aperiodic, as in
Figure~\ref{fig:signals}B, or sinusoidal, as in
Figure~\ref{fig:signals}C.  Rhythmic (sinusoidal or periodic) changes
in the LFP tend to reflect coordinated firing patterns across neurons
in the population, whereas uncoordinated changes are reflected as
changes in the volatility of the LFP~\citep{MannEtal09a, FrieEtal07,
  Buzs06, CronEtal11}.

When both spike timing information and LFP recordings are available,
it is possible to examine whether a given neuron's spikes are
modulated according to the activity of the surrounding population.
For example, \textbf{phase-locked neurons}~\citep{JacoEtal07} tend to
fire action potentials during a particular phase (angle) of an
oscillation that appears in an LFP (Fig.~\ref{fig:signals}D).  Other
neurons exhibit \textbf{phase coding} by changing their preferred
phase according to properties of the stimulus, ongoing experience, or
behavior.  For these neurons, the phases (of LFP oscillations) at
which spikes occur can carry additional information beyond firing rate
alone~\citep{SkagEtal96, HuxtEtal08, KamoEtal98, Lism05}.  One way of
characterizing phase-depending firing is to compute a
\textbf{spike-triggered average} of the LFP preceding and proceeding
each spike (Fig.~\ref{fig:signals}E).  When a neuron exhibits a phase
preference, its spike-triggered average will look like an oscillation
centered at the neuron's preferred phase.

Although oscillations can sometimes be detected visually by examining
a raw LFP recording, a more rigorous approach is to use signal
processing methods to quantify the presence of oscillatory components
of the LFP.  A \textbf{power spectrum} (Fig.~\ref{fig:signals}F) plots
the \textbf{power} at each frequency-- i.e., the extent to which
oscillations at each frequency contribute to the LFP.  When the LFP
exhibits an oscillation, this appears as a peak (centered on the
oscillation's frequency) in the LFP's power spectrum.  An LFP may also
exhibit \textit{multiple} oscillations, which appear as multiple peaks
in the power spectrum.

In addition to true (sinusoidal) oscillations in the LFP, the
volatility of the LFP can also change its power spectrum.  For
example, an increase in the standard deviation of the LFP's changes in
voltages across successive timepoints will result in an increase in
power at \textit{all} frequencies.  So-called \textbf{broadband
  shifts} in power can occur when the neurons in the underlying
population change their firing rates~\citep{MannEtal09a}.

Given the many ways to measure and characterize neural responses,
which approach is best?  The answer depends in part on what we hope to
learn.  For example, if we are interested in processes that we expect
to depend on very precise timing and relatively simple neural
computations, then neuron-centric signals like spike timing and firing
rate may be especially promising.  If we are instead interested in
processes that we expect to depend on large-scale computations carried
out by populations of thousands of neurons, then we may instead
benefit from focusing on periodic and aperiodic features of local
field potentials recorded from relatively large electrodes.  In
general, lower-level processes (e.g., signal transduction) tend to
rely on smaller numbers of neurons and occur over shorter timescales.
Approaches that operate over few neurons and that support high
temporal resolution are often best-suited to studying these low-level
processes.  By contrast, high-level processes (e.g., scene
understanding, complex planning, emotional processing) tend to rely on
large populations of neurons and occur over relatively long
timescales~\citep{BaldEtal17}.  Approaches that record from larger
populations of neurons and that measure processes or changes that
unfold over longer timescales are often best-suited to studying these
high-level processes.  While these general principles have
\textit{tended} to hold across many studies and recording modalities,
it is worth noting some exceptions.  For example, single-neuron
responses in humans~\citep{QuirEtal05} and non-human
primates~\citep{PoldEtal99} can sometimes exhibit selectivity for
high-level stimuli and semantic concepts.

\subsubsection{Units and patterns versus networks.}
\begin{figure}[tp]
\centering
\includegraphics[width=\textwidth]{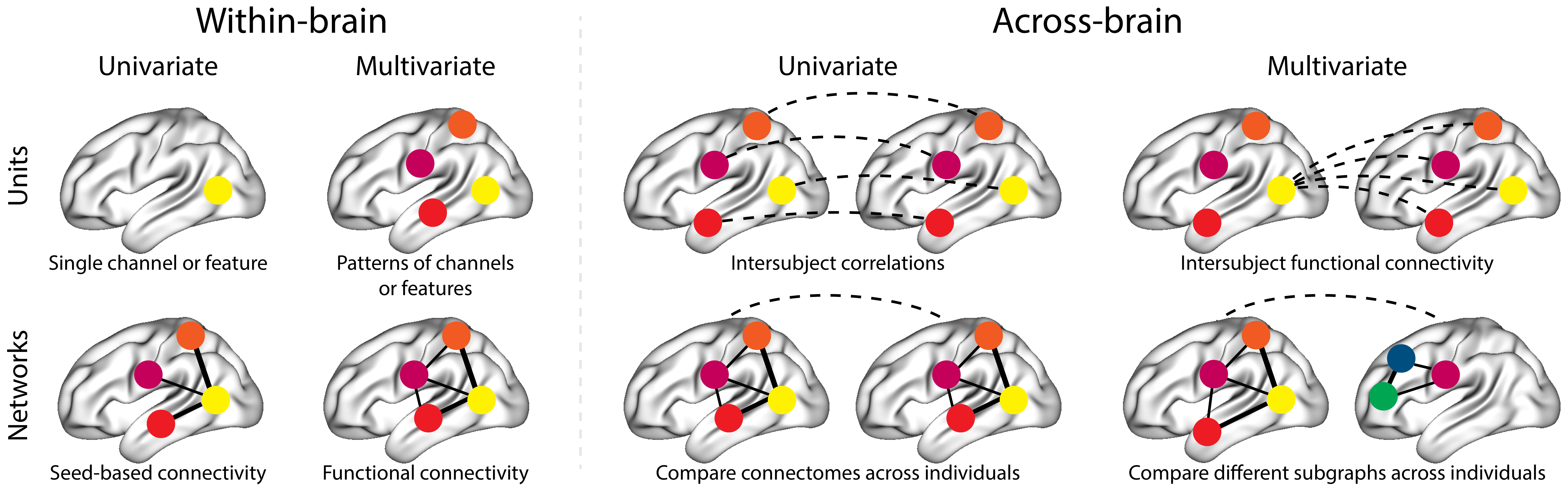}
\caption{\textbf{Univariate and multivariate patterns.}  Individual
  electrodes (channels) or features (e.g., firing rate, phase, power
  at a given frequency, etc.) may be considered in isolation
  (univariate patterns) or in combination with other channels or
  features (multivariate patterns).  Each feature or pattern may be
  considered as an ``independent'' functional unit, or within the
  context of its broader network.  Finally, analyses may be carried
  out on data from a single patient's brain (within-brain) or in
  aggregate across multiple individuals (across-brain).  Note:
  portions of this figure are adapted from~\citep{OwenEtal21}.}
\label{fig:patterns}
\end{figure}

After identifying a set of signals that we think will be appropriate
for studying our phenomenon or cognitive process of interest, the next
key decisions regard whether we should treat those signals in
isolation, or as part of a broader network.  Essentially this comes
down to a decision about how to combine signals and features within
and across participants (Fig.~\ref{fig:patterns}).

Early single-neuron recordings (in cats and non-human primates) played
a central role in Hubel and Wiesel's Nobel Prize-winning work on
mapping out receptive fields of visual cortical
neurons~\citep{HubeWies62, HubeWies68}.  They mapped out the
\textbf{receptive fields} of neurons in the primary visual cortex by
measuring their firing rates as a function of the visual stimulus
shown on the retina.  In general, a neuron's receptive field describes
the stimulus to which it is maximally responsive.  Hubel and Wiesel's
work showed that the primary visual cortex is organized into
\textbf{orientation columns} of neurons whose receptive fields are
tilted dark or light bars at a particular orientation relative to
horizontal.  Several decades later, researchers used high-field fMRI
to show analogous orientation columns in human primary visual
cortex~\citep{YacoEtal08}.  The receptive fields of individual neurons
can be enormously complex.  In contrast to the simple stimuli
preferred by primary sensory neurons, neurons in other brain regions
can have receptive fields that correspond to high-level concepts.  For
example, hippocampal \textbf{place cells} fire preferentially when an
animal travels to a particular location in an
environment~\citep{Tolm48, EkstEtal03, KnieEtal95}.  Other work has
shown that some medial temporal lobe neurons appear to increase their
firing in response to photographs of specific faces, animals, objects,
or scenes~\citep{QuirEtal05}.

When recordings from several neurons are available, the set of firing
rates across the population can provide additional information beyond
that contained in the firing rates of individual
neurons~\citep{PougEtal00, AverEtal06}.  For example, if a single
place cell responds to one area of an environment, a population of
many place cells that each respond to a different location in the
environment can provide a rich cognitive map.

Although they do not provide information about spike timing,
macro-scale LFP recordings also reflect population-level neuronal
activity.  Multi-electrode LFP recordings from one or more brain areas
can be especially informative.  For example, multi-channel LFPs may be
used to decode visual stimuli~\citep{BeliEtal08, BeliEtal10}, auditory
stimuli~\citep{BeliEtal10, SmitEtal13d}, speech
production~\citep{PaslEtal12, HamiEtal21, ProiEtal22}, acute pain onset and
intensity~\citep{ZhanEtal18}, and even semantic
representations~\citep{DezfDali20, MannEtal12}.

Intracranial recordings (of individual neurons, populations of
neurons, and LFPs) may also be considered within the context of the
larger brain networks to which they belong or
contribute~\citep{BassSpor17}.  One set of approaches to
characterizing brain networks is informed by \textbf{graph theory}, a
branch of mathematics concerned with characterizing network
architectures, influence, and membership~\citep{BullSpor09,
  RubiSpor10, HoneEtal07, SporBetz16, SporHone06, BassBull06,
  BetzEtal17b, SizeEtal18, OwenEtal21}.  For example, a timeseries of
recorded responses from multiple channels may be used to infer
functional or causal interactions between the associated neural
populations.  After mapping out a network of pairwise connections
between the responses, graph theoretic measures may be applied to
estimate or compare the influence of a given channel or set of
channels.  Considering interactions can provide information beyond the
responses of individual channels or patterns.  For example, patterns
of interactions between neurons or populations can show selective
modulation in response to stimuli or features, even when the
underlying individual neurons or populations do \textit{not} appear
responsive to the stimulus when considered in
isolation~\citep{RigoEtal13}.

\subsubsection{Static versus dynamic measures of brain activity.}
When we attempt to discover the neural patterns associated with a
particular stimulus or representation, we need to consider two
fundamental questions about how the relevant patterns might change
over time.  The first question is whether brain representations are
fundamentally stable.  For example, each time you think of a concept,
like the meaning of the word ``automobile'' do the brain areas
relevant to representing that concept display the same basic activity
patterns?  Or do the neural representations of concepts change in
meaningful ways over time, such that the representation of a concept
looks fundamentally different each time we measure it?  The second
question is about whether representations themselves are static or
dynamic.  For example, when you think of the concept ``cat,'' does the
entire representation essentially become activated as a single unit?
Or do different components of the representation (e.g., ``fur,''
``mammal,'' ``whiskers,'' ``tail,'' etc.) come online in sequence,
perhaps in a stereotyped way that adds additional nuance or
meaning?

Some of our conceptual knowledge, and presumably the underlying neural
representations of that knowledge, is acquired over timescales on the
order of several years.  For example, as they develop, children
acquire new representations of concepts and how they are related or
organized~\citep{KrolSund03, BlayEtal06}.  Changes in neural
representations that occur over the course of years are unlikely to be
captured by intracranial recordings, which are typically made over
timescales on the order of days or weeks.

Another process that leads to changes in the neural representations is
\textbf{pattern separation}.  Pattern separation refers to the
phenomenon of differentiating the neural representations of two or
more related stimuli or concepts.  For example, pattern separation can
occur when we learn to identify and focus in on subtle differences
between stimuli or concepts that initially seemed (nearly) identical.
Pattern separation can occur over relatively short timescales, and can
be identified using intracranial recordings~\citep{LeutEtal07b,
  BakkEtal08, Roll13, YassStar11}.

Although research on the development of conceptual knowledge and
pattern separation shows that neural representations \textit{can}
change over time, there is also substantial evidence that neural
representations are at least somewhat stable over timescales of hours
to days, and even across different individuals~\citep{MitcEtal08a,
  HaxbEtal11, ShinEtal08}.  For the most part, this body of work
treats the neural representations of concepts and stimulus responses
as essentially static.  By assuming that the neural patterns evoked by
a fixed stimulus are stable (within and across presentations), one can
use a machine learning approach called \textbf{pattern classification}
to learn mappings between neural patterns and stimulus
labels~\citep{NormEtal06b}.  Once these mappings are learned, they may
be applied to new neural patterns to estimate the stimulus or
``thoughts'' associated with those neural patterns.  This allows
researchers to estimate the cognitive dynamics that occur during
neuroimaging~\citep{PolyEtal05a, PolyEtal07, GersEtal13, MannEtal16,
  ChiuEtal21}.

Some stimuli, such as words, images, pure tones, etc., are
perceptually static.  Most or all of the information in the stimulus
is ``made available'' to our sensory systems at the same time.
Certainly it may take some time for higher-order information to unfold
in our minds, for example when we are presented with a complex or
thought-provoking image.  However, those dynamics are driven by
internal processes rather than (directly) by the stimulus itself.
Other stimuli, such as movies, motion sequences, and dynamic sounds
like speech or music, are fundamentally dynamic.  For example, if we
were presented with only the average (across time) visual or auditory
information in a popular feature-length film, we would be missing
nearly all of the structure that made that film engaging or
interesting.  Dynamic stimuli, particularly \textbf{naturalistic
  stimuli} with rich spatiotemporal structure reminiscent of
real-world experiences, can evoke dynamic neural responses that are
often highly reliable across repeated presentations to a single
participant participant, as well as across
participants~\citep{ChenEtal17, KhosEtal21, KiEtal16, NastEtal18,
  HassEtal10, Finn21, ChanEtal20, FinnBand21, MannEtal18, OwenEtal21,
  HuthEtal12, HuthEtal16}.  While it is possible to temporally average
across the timepoints of responses to some classes of dynamic stimuli
while still achieving high reliability~\citep{NastEtal18}, this is not
universally true.  For example, in the domains of speech comprehension
and speech production, temporal information is a primary indicator of
meaning.  Studying neural responses to speech therefore requires
considering how neural correlates of speech unfold over
time~\citep{HamiEtal21, ProiEtal22}.

Independent of whether a given stimulus is fundamentally static or
dynamic, neural responses can also change according to which other
stimuli were experienced nearby in time.  For example, interpreting
the neural response to `B' in the sequence A B C D B might entail
accounting for whether the given instance of `B' is the one that
follows `A' or `D'.  Randomizing stimulus order and averaging over
repeated trials effectively removes this sort of contextual
information.  However, in some cases, the \textbf{context} in which a
stimulus occurs---i.e., the set of other stimuli and thoughts that
were experienced nearby in time---can play a critical role in how we
process, interpret, and remember incoming information~\citep{Mann20}.
For example, priming participants using different cues can reliably
bias them to interpet an ambiguous narrative in a particular
way~\citep{YeshEtal17}.  Accounting for these sorts of contextual
effects often entails factoring in stimulus order or content to the
corresponding analyses and models.

\subsection{Building explicit stimulus models \label{sec:stimmodel}}
Identifying neural responses to a stimulus requires formalizing what the
stimulus in question \textit{is} and when the participant is exposed to
it.  Broadly, this entails building explicit or implicit models that
describe the composition and/or dynamics of the stimulus.  The
features, outputs, or predictions of these models may then be related
to neural patterns.

\subsubsection{What is a stimulus ``model''?}
\begin{figure}[tp]
\centering
\includegraphics[width=\textwidth]{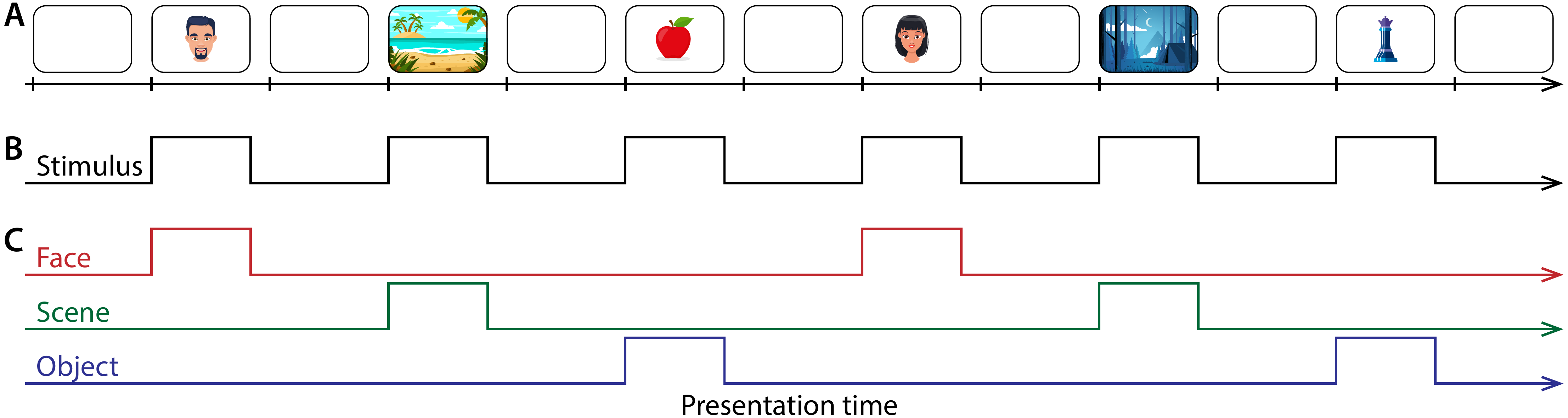}
\caption{\textbf{Discrete stimulus timeseries.}  \textbf{A.  Example
    stimulus sequence.}  A succession of images are presented to the
  participant on a computer screen, interspersed by intervals of blank screen.  Images are
  drawn from three categories: faces, outdoor scenes, and concrete
  objects.  \textbf{B.  Onset and offset timing.}  Stimulus timing,
  but not stimulus category, may be conveyed using a single binary
  timeseries.  \textbf{C. Stimulus identity.}  Stimulus identity
  (e.g., category) may be represented using a single binary timeseries
  for each stimulus category or feature.}
\label{fig:binary}
\end{figure}

From an analytic perspective, describing a stimulus typically entails
characterizing how different aspects of the stimulus change over
time.  If we are solely interested in the presence or absence of a
stimulus, or the timings of a discrete sequence of trials (Fig.~\ref{fig:binary}A), then the
``stimulus'' might be describable as a simple binary sequence
(Fig.~\ref{fig:binary}B).  One could then examine or compare neural
activity patterns recorded during the ``on'' timepoints versus ``off''
timepoints, or estimate how neural responses change during the
transitions between those binary states.  If stimuli are drawn from a
well-defined set of categories, then category information may be conveyed
using one binary sequence per category (Fig.~\ref{fig:binary}C).

\begin{figure}[tp]
\centering
\includegraphics[width=\textwidth]{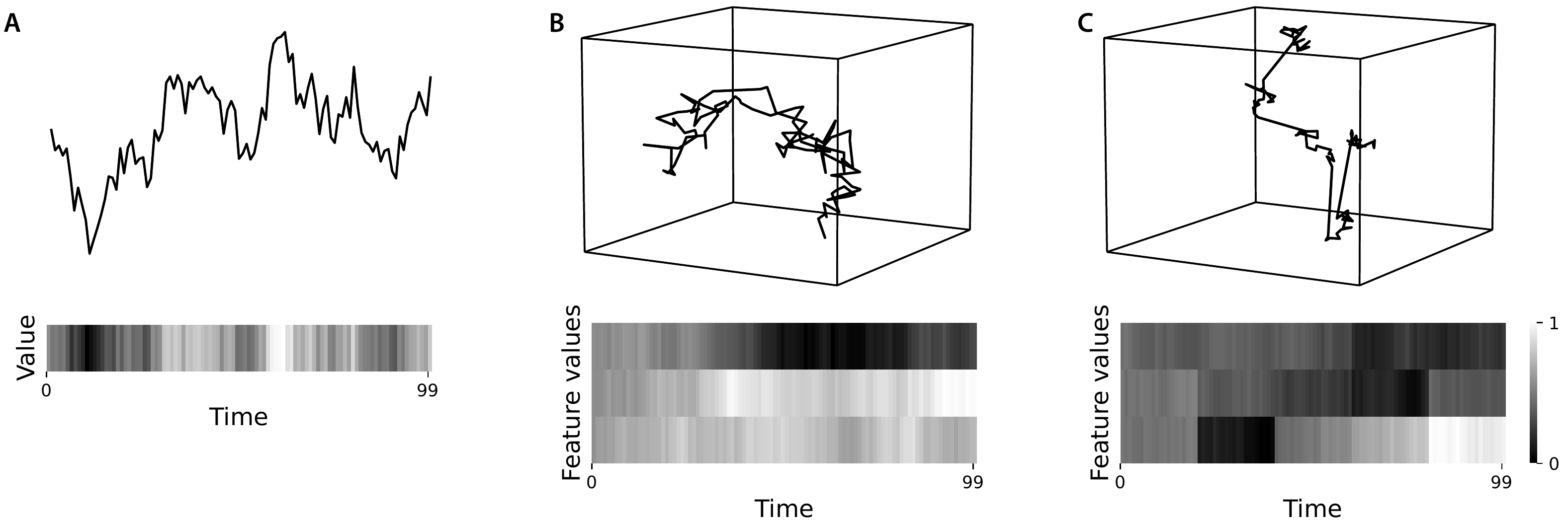}
\caption{\textbf{Continuous stimulus timeseries.}  
  \textbf{A. Univariate real-valued stimulus.}  The stimulus takes on
  any value at any timepoint.  In this example the stimulus values are
  autocorrelated.  \textbf{B. Multivariate real-valued stimulus.}  The
  stimulus comprises multiple features, each of which can take on any
  value at any moment.  When projected into 3D, the stimulus traces
  out a \textbf{trajectory} describing how the values of its features
  change over time.  \textbf{C. Multivariate real-valued stimulus with
    event-level or trial-level dynamics.}  This stimulus is similar to
  the one displayed in Panel B, but here the stimulus values exhibit
  occasional event boundaries.}
\label{fig:stimuli}
\end{figure}

In other instances, we may be interested in understanding how neural
responses relate to specific stimulus values, or how those values
change over time.  For example, we might describe the brightness or
salience of a visual stimulus (or the loudness of an auditory
stimulus, etc.) as a sequence of real-valued numbers
(Fig.~\ref{fig:stimuli}A).  This could enable us to understand graded
neural responses to the stimulus, such as how neural responses change
as a function of the stimulus values.  \textbf{Event-triggered averages}
(analogous to spike-triggered averages such as
Fig.~\ref{fig:signals}E) can also provide insights into how the
stimulus tended to change during a time window centered on a
particular neural event (e.g., an action potential or the appearance
of a specific activity pattern).

Some stimuli are best described by multivariate real-valued feature
  vectors whose elements (i.e., ``features'') describe the absense,
presence, or values of specific stimulus properties
(Fig.~\ref{fig:stimuli}B).  Describing a stimulus as a timeseries of
multivariate feature vectors can facilitate more nuanced mappings
between those stimulus properties and different aspects of neural
activity.  For example, the firing rates of different neurons, or the
patterns of power spectra across the electrodes in a given region of
interest, might display different sensitivity to different stimulus
features.

A fourth (general) way of describing how a stimulus changes over time
is to use a multivariate timeseries with explicit event-level or
trial-level boundaries (Fig.~\ref{fig:stimuli}C).  For example, in a
movie, the scene cuts could constitute \textbf{event boundaries}--
i.e., moments of transition where the stimulus features exhibit rapid
``jumps'' that are substantially larger than usual between-timepoint
changes.  Event boundaries can delineate changes in the focus of an
ongoing conversation, scenes in a story or movie, environmental
changes, or other transitions in the low-level or high-level content
of the stimulus.  Experimental trials can also be considered as a sort
of event boundary.  For example, a multivariate timeseries like that
in Figure~\ref{fig:stimuli}C could also be used to describe the
content of a sequence of short video clips presented in succession.
Within each clip, the features might change comparatively less than
across clips.

As described next, there are many ways to define what the stimulus
features \textit{are}.  This requires making assumptions about which
aspects of the stimulus ``matter'' (e.g., in terms of evoking neural
responses, predicting behaviors, etc.), and about how the moments of
the stimulus timecourse should be matched up with moments of a neural
recording.  While these assumptions can have a large influence on the
outcome of an analysis, there is unfortunately no universal way of
describing or modeling stimuli (or of relating stimulus dynamics to
neural responses).  Rather, one must make informed decisions about how
to procede based on the sorts of insights one hopes to gain from an
analysis, or based on what approaches have performed well in prior
related work. 

\subsubsection{Manual approaches.}
In trial-based experiments where stimuli (e.g., words, sounds, images,
etc.) have well-defined onset and offset times, the stimulus onset and
offset timeseries (e.g., Fig.~\ref{fig:binary}B) can serve as a
simple stimulus ``model.''  A binary sequence that solely describes
stimulus onset and offset times ignores \textbf{stimulus identity}
(e.g., the category or label) and \textbf{stimulus features} (e.g.,
the values of the corresponding feature vectors).  In this way,
modeling the stimulus using a binary sequence makes the implicit
assumptions that (a) stimulus timing is the main factor of interest with
respect to the associated neural responses and (b) stimulus identity
and stimulus features may be safely ignored.

To model stimulus identity \textit{and} timing (when both identity and
timing information are well-defined, e.g. as in trial-based
experiments), the stimulus may also be modeled using a \textit{set} of
binary timeseries (Fig.~\ref{fig:binary}C).  For example, one might
define a separate binary timeseries describing the onsets and offsets
of only one stimulus category or trial type.  This approach makes the
implicit assumption that the specific identities of different
exemplars (e.g., within a stimulus category-- such as face images of
different people's faces) may be safely ignored in favor of
prioritizing coarser-scale information such as broad stimulus category
or trial type labels.

When parameterized stimuli are constructed to vary along one or more
explicit stimulus dimensions (e.g., visual stimuli that vary in
brightness, contrast, spatial frequency, etc.), the values along each
dimension and/or the parameters themselves can serve as a
representation of the stimulus (e.g., Fig.~\ref{fig:stimuli}).  Each
stimulus dimension (or parameter) may be represented by its own
real-valued timeseries.  This approach makes the implicit assumption
that the only relevant sources of variation in the stimulus are those
characterized by the specified stimulus dimensions or parameters.  All
other stimulus features or properties are effectively ignored.

Some stimuli cannot be adequately characterized or described using
their associated parameters and/or features that can be directly
mapped onto specific physical or perceptual properties.  For example,
such stimuli might be best described using high-level perceptual or
conceptual properties of the stimulus such as the presence of specific
objects or high-level content, emotional tone, etc.  When these
properties may be readily judged or rated by human observers, normed
ratings or judgements (typically collected by an independent set of
participants) may be used as another means of quantifying stimulus
features.  These judgements may take the form of integer-valued or
real-valued responses (e.g., rating how ``happy'' an image or sound
is, on a particular scale) or binary ``yes/no'' judgements (e.g.,
indicating whether or not a tree is present in an image).

\subsubsection{Automated approaches.}
Many types of stimuli, including natural images, text, complex sounds
(e.g., speech, music, recordings of natural environments, etc.),
movies, and others, cannot always be easily categorized or manually
labeled or rated.  In other cases, even if manually labeling stimuli
might be possible in principle, in practice it may be too expensive in
money or time to generate manual labels.  Automated approaches to
building stimulus models can scale to millions of stimuli and
thousands of stimulus features.

One class of automated approaches to generating stimulus feature
models entails applying probabilistic models and deep neural networks
such as convolutional neural networks~\citep{LeCuBeng95, KrizEtal12},
text embedding models~\citep{DeerEtal90, LandDuma97, NelsEtal04,
  SteyEtal04, BleiEtal03,
  BengEtal03, BrowEtal92, MnihHint09, MikoEtal13a, MikoEtal13b,
  PennEtal14, PeteEtal18, KiroEtal15, CerEtal18, ConnEtal18,
  WietKiel19}, transformers~\citep{DevlEtal18, BrowEtal20,
  RadfEtal19}, and others~\citep{KharEtal21} to the stimuli.  The
activations of the hidden or output layers of these networks may be
used as feature vectors for the corresponding stimuli.

Another class of automated approaches includes visual entity
tagging~\citep{ChenEtal13b}, image-to-text models~\citep{StonEtal18,
  BaboEtal11, deAnEtal18}, speech-to-text models~\citep{FuruEtal04,
  JoneEtal03, ZimaEtal18, WilpEtal90}, and other algorithms for
generating text data from non-text stimuli like still images, video,
and sound.  After training these models on large corpora of labeled
examples, novel stimuli may be automatically tagged with text
annotations.  In turn, these annotations may be passed onto text
embedding models to construct feature vectors for each stimulus.
Alternatively, the annotations may be treated directly as stimulus
features, for example by treating each unique keyword as a binary
feature that is either present or absent in each stimulus.

\textbf{Human-in-the-loop} techniques~\citep{Zanz19} provide a balance
between purely manual and purely automated methods.  These techniques
entail combining human feedback with classic machine learning
approaches.  For example, \textbf{multidimensional
  scaling}~\citep{EnniEtal88, CarrChan70, Torg58} may be applied to
pairwise similarity judgements from human participants to derive
$n$-dimensional feature vectors whose pairwise correlations or
(inverse) distances are consistent with those
judgements.

To facilitate comparisons or other analyses, it can be useful to
partition continuous stimuli into discrete ``states'' or ``events''.
For example, \textbf{hidden Markov models} may be applied to a
multidimensional timeseries of observations (e.g., feature vectors) to
estimate the moments of transition that are interspersed between
periods of relative stability~\citep{Rabi89, BaldEtal17, HeusEtal21}.
Neural responses during different events, or at transition points, may
then be examined further.

\subsection{What are some modality-specific challenges to identifying
  stimulus-driven brain activity from intracranial recordings?}

\begin{figure}[tp]
\centering
\includegraphics[width=0.5\textwidth]{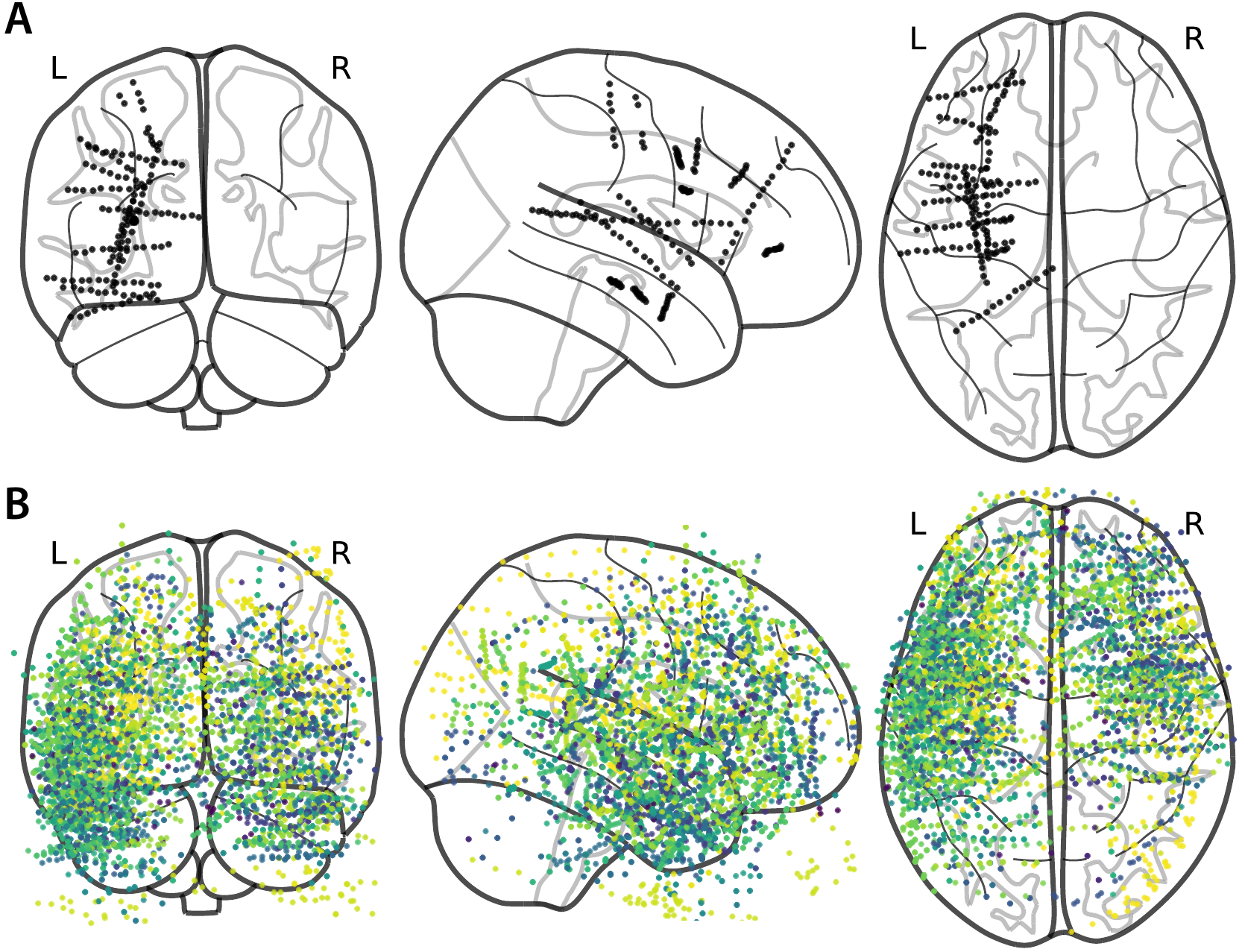}
\caption{\textbf{Within-patient versus across-patient electrode coverage.}
  Each dot denotes the location of the recording surface of one
  neurosurgically implanted electrode.  \textbf{A.  Example patient.}
  The locations of the 169 electrodes implanted in one patient's brain
  are displayed.  \textbf{B. Across-patient electrode locations.}  The
  locations of $n = 5023$ electrodes are displayed.  Colors denote
  different patients (electrodes from $m = 53$ patients are
  displayed).  Both panels: implantation locations are taken from \cite{EzzyEtal17}
  and filtered using a thresholding procedure to remove noisy signals
  reported by \cite{OwenEtal20}.}
\label{fig:electrodes}
\end{figure}

While intracranial recordings provide high spatiotemporal resolution
data about neural activity (Fig.~\ref{fig:spacetime}), the
\textbf{coverage} afforded by intracranial recordings---i.e., the
proportion of the brain volume captured across all electrodes---is
relatively poor compared with popular non-invasive approaches like
fMRI, MEG, and scalp EEG.  The locations of electrodes implanted in a
representative patient's brain are shown in
Figure~\ref{fig:electrodes}A.  Across the 169 electrodes, coverage is
limited primarily to the left frontal and temporal lobes.  The
recordings from this example patient cannot provide direct information
about activity outside of these regions.

The precise electrode numbers and implantation locations are
determined by teams of clinicians whose primary goal is (typically) to
locate the seizure focus for that patient.  For some patients, the
clinical team may have a relatively good sense of where the patient's
seizures likely originate.  These patients may be implanted with relatively few
electrodes, spread over a relatively small area of the brain.  For
other patients where the seizure focus is less clear, they may be
implanted with a larger number of electrodes spread throughout the
brain.  Taken together, these factors mean that each patient is
implanted with a different number of electrodes, in different
locations, according to a unique clinical plan.  Whereas non-invasive
recordings can often be easily aligned across people (since the
sensor locations are typically held relatively constant across
people), the alignment problem is highly non-trivial for intracranial
recordings.

One benefit to having varied electrode locations across patients is
that, for a sufficiently large number of patients, it becomes possible
to obtain data from most of the brain (Fig.~\ref{fig:electrodes}B).
Full-brain maps and analyses may be obtained from intracranial
recordings by constructing maps that are stitched together across
patients.  These maps obtained from intracranial recordings may then
be compared to analogous maps obtained using other neuroimaging
approaches to provide clues about the reliability of the
across-patient findings.  However, because full-brain maps derived
from intracranial recordings typically require combining data across
patients, it can be difficult to identify reliable within-patient
effects, or to compare responses across patients.

\section{Identifying stimulus-driven neural activity \label{sec:linking}}
Thus far, we have surveyed a variety of approaches for measuring or
characterizing neural activity patterns and stimuli.  When applied in
conjunction, the result of these approaches is a set of two
timeseries: one describing the patient's neural responses and the other describing
the stimulus the patient experienced as they exhibited those neural
responses.  The final step is to combine these characterizations in
order to relate changes in neural activity to changes in the
stimulus.  Broadly, this combination step may be carried out
within participant or across participant.

\subsection{Within-participant approaches}
\textbf{Within-participant} analyses are carried out on data from a
single person.  Identifying stimulus-driven neural activity using
within-participant analyses typically entails combining data over time
(e.g., across runs, conditions, trials, etc.).  The objective is to
estimate maps, patterns, or response profiles that are unique to each
individual.  These within-participant estimates may then be combined
across participants to examine general tendencies in the population and/or
individual-specific markers.

\subsubsection{Generalized linear models and multivariate pattern
  analysis.}

Given a neural recordings (Sec.~\ref{sec:activity}) and a stimulus
model (Sec.~\ref{sec:stimmodel}), the two most widely used approaches
to identifying stimulus-driven neural activity are generalized linear
models and multivariate pattern analysis.  Broadly,
\textbf{generalized linear models} (GLMs) are an approach to
predicting a set of labels, $\mathbf{Y}$, typically represented by one
or more feature vectors, from a set of an equal number of inputed
observations, also represented by feature vectors,
$\mathbf{X}$~\citep{NeldWedd72}.  Formally, we say that
\[
y_t = f\left(x_t, \beta\right),
\]
where $y_t$ is the $N$-dimensional vector of output labels for
observation $t$, $x_t$ is a the $M$-dimensional
$t$\textsuperscript{th} observation, and $\beta$ is an $M \times N$
matrix of weights.  The \textbf{link function}, $f(\cdot)$ takes as
inputs a set of observations and weights and produces as output a
transformed version of the $N$-dimensional vector $\beta^\top x_t$.
For example, if the elements $\beta^\top x_t$ are Real-valued and lie
within the interval $(-\infty, +\infty)$, then a sigmoidal link
function (e.g., the logistic or hyperbolic tangent functions) would
transform the elements to lie within the interval $(-1, +1)$.  The
``power'' of the generalized linear model framework comes from the
flexibility in how the link function may be defined.  By choosing an
appropriate link function, it is possible to take arbitrary
Real-valued inputs and transform them into outputs that match a wide
variety of useful formats-- e.g., (unbounded) Real-valued outputs;
probability-like values bounded between 0 and 1; indicator vectors
(i.e., vectors where all values are 0s except for one element whose
value is 1); binary-like values (i.e., where extreme values are
``pulled'' towards one of two boundaries, as in sigmoid functions);
and many more.  When $\mathbf{X}$ reflects neural activity and
$\mathbf{Y}$ reflects the stimulus features during the corresponding
moments, the fitted GLM weights (i.e., $\beta$) describe how different
aspects of neural activity relate to different stimulus features.
These fitted weights may also be used to ``decode'' stimulus features
from new, previously unobserved, neural data.

\textbf{Multivariate pattern analysis} (MVPA) describes a second class
of approaches for connecting stimulus features and neural
activity~\citep{HaxbEtal01, NormEtal06b}.  Like GLMs, the goal of MVPA
is to predict a set of labels from a set of observations.  Typically
the ``labels'' comprise stimulus features and the ``observations''
comprise neural responses.  GLMs are a special case of MVPA for which
the output features reflect a (potentially transformed) linear
combination of the input features.  However, MVPA also includes a
variety of other approaches for which the relations between input and
output features are non-linear, and potentially even non-monotonic.
The umbrella term for such algorithms is \textbf{pattern classifiers},
and includes GLMs, support vector machines~\citep{BoseEtal92},
boosting~\citep{Scha03}, naive Bayes classifiers~\citep{CortVapn95,
  Nobl04, LeeEtal12}, nearest neighbor-based
classifiers~\citep{FixHodg51}, and (deep) neural network-based
classifiers~\citep{ZengEtal14, Wan90}, among others.

\subsubsection{Representational similarity analysis.}
Comparing the \textbf{neural temporal correlation matrix} (i.e., the
correlations between the neural patterns recorded at every pair of
timepoints) to the \textbf{stimulus temporal correlation matrix}
(i.e., the correlations between the stimulus feature vectors at every
pair of timepoints) can reveal similarities and differences between
how neural and stimulus feature change over time.
\textbf{Representational similarity analysis} (RSA) entails computing
the element-wise correlation between the upper triangles of the neural
and stimulus correlation matrices~\citep{KrieEtal08b}.  Following the
logic of \cite{KrieEtal08b}, to the extent that neural patterns show a
similar temporal correlation structure to stimulus features, we can
interpret this as evidence that the neural and stimulus features are
related.

RSA may also be carried out across a series of \textbf{searchlights}.
Similar to how a searchlight illuminates a well-defined area of
darkness, a searchlight analysis provides insights into the
functionality or responses profiles of a focused region
of interest.  For each of a series of spherical volumes tiled
throughout the brain, RSA may be performed for each volume by limiting
the neural features under consideration to only those captured by
electrodes within that sphere's radius.  This yields, for each sphere
(i.e., searchlight) a single correlation coefficient between that
sphere's neural temporal correlation matrix and the stimulus temporal
correlation matrix.  Examining which searchlights displayed high
versus low correlations can highlight which brain areas might
represent the stimulus in a way consistent with a given stimulus
model (i.e., the model used to construct or estimate the stimulus
features).

A convenient property of RSA is that, unlike approaches like GLMs or
MVPA, RSA does not require learning an explicit mapping between neural
features and stimulus features.  This is because RSA is driven solely
by pattern similarities across timepoints, rather than by the specific
properties of the patterns themselves.  In this way, RSA can sometimes
be a more sensitive way of identifying stimulus-driven neural patterns
(e.g., compared with GLMs and MVPA).  For example, high levels of
noise during many of the exposures to a particular stimulus category
will mean that neural decoding approaches will likely fail to
effectively learn mappings between the neural and stimulus features
for that category.  However, RSA analyses effectively ``average'' over
all timepoints, thereby highlighting aspects of neural activity with a
similar temporal correlation structure to \textit{any} stimulus
features (even if the associations with a subset of the stimulus
features are noisy).

\subsubsection{Joint stimulus-activity models.}
Thus far, we have reviewed two approaches to identifying
stimulus-driven neural activity.  MVPA attempts to learn mappings
between stimulus features and neural features, and RSA attempts to
identify stimulus features and neural features that exhibit similar
temporal correlation patterns.  A third (related) approach entails
building models that jointly consider the timecourses of neural and
stimulus features.  Whereas MVPA and RSA implicitly treat stimulus and
neural features as a ``ground truth,'' joint stimulus-activity models
allow the stimulus features and neural features to mutually inform
each other.  These models assume that, while the \textit{mappings}
between stimulus and neural features may be relatively stable, there
may be some times when the stimulus features provide a more reliable
signal and other times when neural features provide a more reliable
signal~\citep{TurnEtal13}.

\begin{figure}[tp]
\centering
\includegraphics[width=\textwidth]{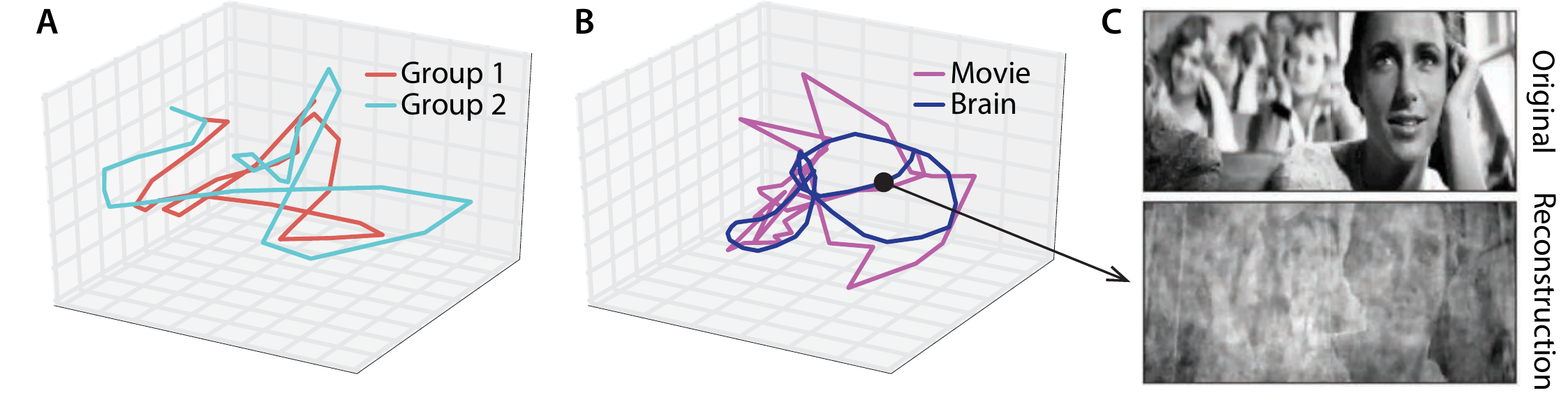}
\caption{\textbf{Joint geometric models of stimulus and neural
    features.}  \textbf{A.  Neural features from different
    individuals.} Group-averaged trajectory of fMRI activity from
  ventral visual cortex split into two randomly-selected groups of
  subjects (group 1: $n=6$, group 2: $n=5$) watching the same
  movie~\citep{HaxbEtal11}. \textbf{B.  A common geometric space for
    stimulus and neural features.} Group-averaged trajectory of fMRI
  activity from ventral visual cortex and trajectory of movie (pixel
  intensities over time) hyperaligned~\citep{HaxbEtal11} to a common
  space.  In Panels A and B, the high-dimensional stimulus and neural
  trajectories have been projected onto three dimensions to facilitate
  visualization~\citep{HeusEtal18a}.  \textbf{C.  Interpreting
    coordinates in the common feature space.} Stimuli (e.g., movie
  frames) may be reconstructed from ventral visual brain activity by
  mapping the coordinates of neural features in the common space onto
  coordinates in the stimulus space.  Note: this figure is adapted
  from~\citep{HeusEtal18a}.}
\label{fig:geometry}
\end{figure}

Figure~\ref{fig:geometry} provides some geometric intuitions for the
idea of joint stimulus-activity models.  First, consider how we might go
about estimating the reliability of some neural responses we measure
as people experience a given stimulus, such as watching a movie.  One
approach might be to expose a single participant to the same stimulus
multiple times.  Or alternatively, we might expose several different people to
the same stimulus.  We could then examine the similarities and
differences between the neural responses recorded across trials or
individuals.

Geometrically, the neural features recorded from one individual,
during one moment, can be conceptualized as a single point in a high
dimensional feature space (whose dimensions each correspond to a
single neural feature).  Over time, as the individual's neural
activity changes, the successive activity patterns trace out a
\textbf{trajectory} through this neural activity space.  When the
timecourses of neural activity patterns are similar across trials or
individuals, this results in similar neural trajectory shapes
(Fig.~\ref{fig:geometry}A).

The stimulus features from a single moment, as well as the timecourse
describing how stimulus features change over time, may also be
conceptualized as a trajectory through a (different) high dimensional
feature space, whose dimensions each correspond to a single stimulus
feature.  It could be interesting to ask whether the neural and
stimulus trajectories are similarly shaped, or whether there are
particular moments or circumstances under which the shapes converge or
diverge.  However, because the dimensions of neural trajectories and
stimulus trajectories do not (typically) match, an additional step is
needed before such comparisons may be made.  The \textbf{procrustean
  transformation} is a geometric transformation for bringing
two sets of coordinates into an optimal point-by-point alignment.
This entails computing the affine transformations
(i.e., rotations, reflections, and scalings) that, when applied to
coordinates in one set, minimize the average Euclidean distance
between the corresponding coordinates in the second set.  The
resulting aligned coordinates may then be directly compared, since the
transformations map the coordinates in the first set into the same
coordinate system (with the same dimensions) as the second
set~\citep{HaxbEtal11}.  Mapping a neural trajectory into the stimulus
feature space (Fig.~\ref{fig:geometry}B, C) provides a common coordinate
system for describing both stimulus features and neural features.

We can use this geometric framework to conceptualize what it means to
jointly model the stimulus and neural responses.  Consider, for
example, the stimulus and neural trajectories displayed in
Figure~\ref{fig:geometry}B.  While both trajectories look similar in
some respects (e.g., they have roughly similar coarse-scale shapes),
they also differ in potentially important ways (e.g., the stimulus
trajectory is ``spikier'' than the neural trajectory).  Which
trajectory is ``correct''?  On one hand, the stimulus trajectory
provides a relatively clean characterization of the stimulus that
exactly reflects specific measurable aspects of what participants were
exposed to.  In this sense, the stimulus trajectory is not
``corrupted'' by measurement noise, inattention, or other factors
unrelated to the stimulus itself.  On the other hand, the stimulus
trajectory is (by definition) a reflection only of the specific
stimulus features that we, as the experimentalists, decided were
likely to be important.  Those features might be at best different and
at worst unrelated to the stimulus properties that participants
actually care about or respond to.  In this sense, one could argue
that neural responses reflect the most direct representation of
aspects of the stimulus the brain is responding to, since those neural
responses are uncorrupted by the experimentalists' assumptions about
which stimulus features are important.  Taken together, it is clear
that neither the stimulus trajectory nor the neural response
trajectory, in isolation, provide a complete reflection of an
individual's internal mental representations of the stimulus.
Instead, it might be most accurate to incorporate aspects of both the
stimulus and neural trajectories.  This joint stimulus-activity
modeling approach acknowledges that the true representation(s) to
which an individual's brain is responding may lie somewhere between
the stimulus and neural trajectories.

\subsection{Across-participant approaches}
\textbf{Across-participant} analyses are carried out on data from
multiple individuals.  Identifying stimulus-driven neural activity
using across-participant analyses typically entails building an
across-participant model or developing analyses that characterize
similarities or differences in responses across participants.  The
objective is to estimate a single map, pattern, or response profile
that is common across individuals.  Some approaches also attempt to
estimate individual differences that characterize how each
individual's responses differ from the group's (aggregated) responses.

\subsubsection{Across-participant models.}
Building across-participant models for identifying stimulus-driven
neural activity requires defining a common representation for
describing neural activity (and, potentially, linking neural features
with stimulus features).  Each participant's data must first be mapped
into the common representation space.  This may be carried out using
anatomical~\citep{HermEtal10, ArchEtal07, StudEtal00},
functional~\citep{HaxbEtal11, ChenEtal21, XieEtal21, HaxbEtal20}, or
other~\citep{SabuEtal10, SekiEtal02} alignment methods.  Next, the
\textbf{inference procedure} (i.e., the algorithm for estimating model
parameters from the observed data) must learn both \textbf{local
  parameters} (i.e., parameters that are specific to each individual)
and \textbf{global parameters} (i.e., parameters that are shared
across individuals).  The final step is often to learn a mapping or
linking function for connecting local and/or global parameters to
stimulus features.  This last step can be carried out
within-participant (by learning mappings between local parameters and
stimulus features) or across-participant (by learning mappings between
global parameters and stimulus features).

\paragraph{Hierarchical matrix factorization models.}
\textbf{Matrix factorization} encompasses a family of mathematical
approaches for decomposing a matrix into the product of several other
matrices.  This family includes a large number of machine learning
models, including Topographic Factor Analysis (TFA)~\cite{MannEtal14b,
  MannEtal18}, Topographic Latent Source Analysis
(TLSA)~\cite{GersEtal11}, Principal Components Analysis
(PCA)~\cite{Pear01}, Exploratory Factor Analysis (EFA)~\cite{Spea04},
and Independent Components Analysis (ICA)~\cite{JuttHera91,
  ComoEtal91}, among others.  Within the domain of neuroimaging, the
general formulation is to first organize the neural feature vectors
(from a single subject) into a $T$ by $N$ data matrix, $\mathbf{Y}$
(where $T$ is the number of observations and $N$ is the number of
neural features).  We can then decompose $\mathbf{Y}$ as follows:
\[
  \mathbf{Y} \approx \mathbf{W}\mathbf{F},
  \]
  where $\mathbf{W}$ is a $T$ by $K$ \textbf{weight matrix} (which
  describes how each of $K$ factors are activated for each
  observation) and $\mathbf{F}$ is a $K$ by $N$ matrix of factor
  images (which describes how each factor maps onto the neural
  features).  In the general case, there are infinitely many solutions
  for this decomposition.  Different matrix factorization approaches
  converge on specific choices for $\mathbf{W}$ and $\mathbf{F}$ by
  placing different constraints on the forms the matrices must take or
  by choosing optimization metrics that emphasize different aspects of
  $\mathbf{Y}$ to be preserved.  For example, when $K << N$, the
  approximation of $\mathbf{Y}$ via $\mathbf{W}\mathbf{F}$ will be
  inexact.
  
\begin{figure}[tp]
\centering
\includegraphics[width=0.6\textwidth]{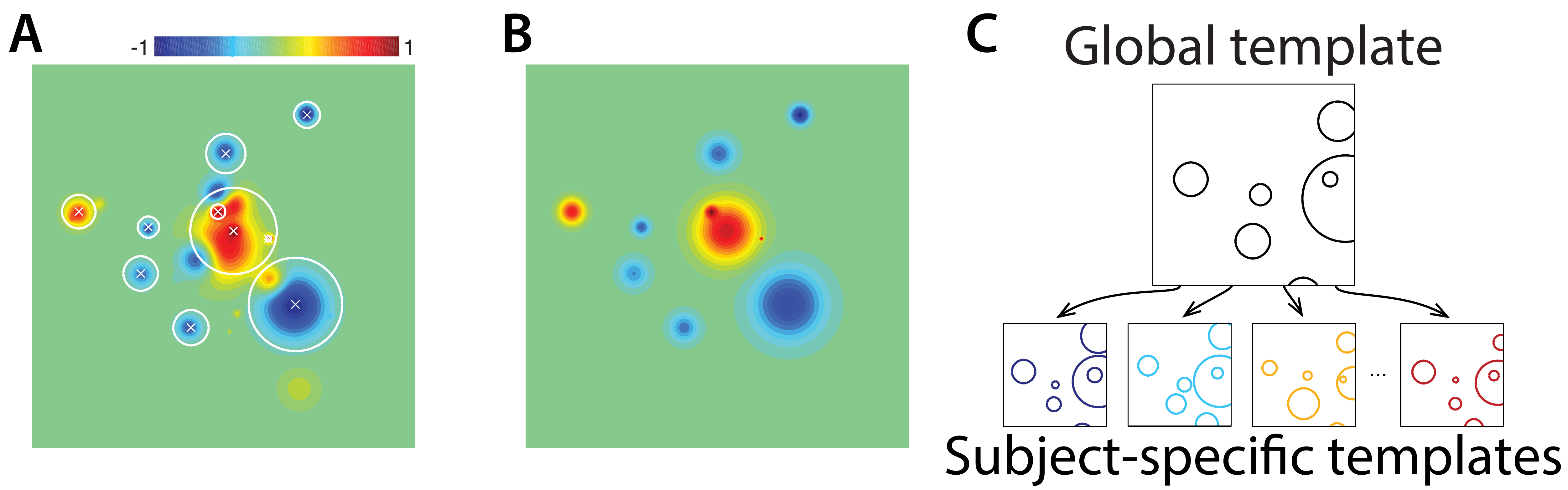}
\caption{\textbf{Topographic Factor Analysis.  A. Spherical factors
    describe contiguous regions of similar activity.}  Each factor is
  represented as a radial basis function.  A factor's image may be
  constructed by evaluating its radial basis function anywhere within
  the brain volume.  Level curves for several example factors fit to a
  synthetic 3D image are outlined in white; $\times$s denote the factor
  centers projected onto the 2D slice displayed in the panel.
  \textbf{B. Brain images are described by weighted sums of the factors'
    images.}  After computing each factor's image (using its radial
  basis function), arbitrary brain images may be approximated using
  weighted combinations of the images for each factor.  The per-image
  weights may be used as a low-dimensional embedding of the original
  data.  A 2D slice of the reconstruction for the image displayed in
  panel A demonstrates how contiguous clusters of locations are
  approximated using weighted activations of spherical
  factors. \textbf{C. The global template serves as a prior for
    subject-specific parameters.}  The global template defines the
  numbers of factors, their locations, and their sizes, for the
  prototypical participant.  Each individual participant's parameters
  (factor locations and sizes) are fit using the global template as a
  prior.  This provides a linking function between different
  participants' factors, thereby enabling across-subject comparisons. A
  subset of the factors outlined in Panel A are displayed in the
  \textit{global template} cartoon.  The positions of these factors in
  each individual participant's \textit{subject-specific template} are
  displayed in different colors.  Note: this figure is adapted
  from~\citep{KumaEtal21}.}
\label{fig:tfa}
\end{figure}

To illustrate how matrix factorization models may be constructed to
capture multi-subject data, we can examine the details of two related
models: hierarchical and non-hierarchical variants of the same matrix
factorization model, topographic factor analysis.  In its
non-hierarchical framing, TFA specifies that each row of $\mathbf{F}$
is parameterized by the center parameter, $\mu$, and the width
parameter, $\lambda$, of a \textbf{radial basis function}.  If a
radial basis function has center $\mu$ and width $\lambda$, then its
activity $\mathrm{RBF}(\mathbf{r}|\mu,\lambda)$ at location
$\mathbf{r}$ is:
\[
  \mathrm{RBF}(\mathbf{r}|\mathbf{\mu},\lambda) =
  \mathrm{exp}\left\{ -\frac{||\mathbf{r} -
   \mathbf{\mu}||^2}{\lambda} \right\}.
\]
The factor images are filled in by evaluating each radial basis
function, defined by the corresponding parameters for each factor, at
the location(s) of each electrode or brain region of interest.  In
contrast to the factors obtained using PCA or ICA, TFA's more
constrained factors may be represented much more compactly; each
factor corresponds to the structure or group of structures in the
brain over which the factor spreads its mass (which is governed by
$\mu$ and $\lambda$).  TFA's factors may be conceptualized as nodes
located in 3D space whose activity patterns influence the observed
brain data (Fig.~\ref{fig:tfa}A, B).

Hierarchical Topographic Factor Analysis (HTFA) works similarly to
TFA, but places an additional constraint over the factors to bias all
of the subjects to exhibit similar factors.  Whereas TFA attempts to
find the factors that best explain an individual subject's data, HTFA
also attempts to find the factors that are common across a group of
subjects (Fig.~\ref{fig:tfa}C).  This is important, because it allows
the model to jointly consider data from multiple subjects.

HTFA handles multi-subject data by defining a \textit{global
  template}, which describes in general where each radial basis
function is placed, how wide it is, and how active its node tends to
be.  In addition to estimating how factors look and behave in general
(across subjects), HTFA also estimates each individual's
\textit{subject-specific template}, which describes each subject's
particular instantiations of each radial basis function (i.e.\ that
subject's radial basis function locations and widths) and the factor
weights (i.e.\ the activities of each of that subject's radial basis
function factors in each of that subject's observed neural activity
patterns).  Because the subject-specific templates are related to each
other (hierarchically~\cite{GelmHill07}, via the global template), a
given factor's radial basis function will tend to be located in about
the same location, and be about as large, across all of the
subject-specific templates.  Because each subject has the same set of
factors (albeit in slightly different locations and with slightly
different sizes) we can run analyses that relate the factors across
subjects.

The general approach of learning global and subject-specific factors
may be applied to many matrix factorization models.  For example,
TLSA~\citep{GersEtal11}, PCA~\citep{WestEtal98}, and
ICA~\citep{WangGuo19} each have hierarchical framings as well.  The
particular benefit of using (H)TFA to decompose and describe
intracranial data is that the radial basis function factors may be
evaluated at the unique locations of each individual patient's
electrodes, even though the electrode locations will differ across
patients.  And because the subject-specific templates are associated
via the global template, aspects of the subject-specific templates may
be compared across patients.  For example, after using HTFA to learn
the global and subject-specific templates, these templates may then be
treated as neural features and examined in relation to stimulus
features.  Specifically, each patient's $\mathbf{W}$ matrix may be
treated as neural features-- but whereas the ``raw'' neural features
in the original dataset will not be consistent across patients, the
columns of each patient's $\mathbf{W}$ matrix may be directly combined
or compared.  In addition, the global and subject-specific factors
(rows of $\mathbf{F}$) may be examined to identify how the columns of
$\mathbf{W}$ map onto different brain areas or structures.

\begin{figure}[tp]
  \centering \includegraphics[width=\textwidth]{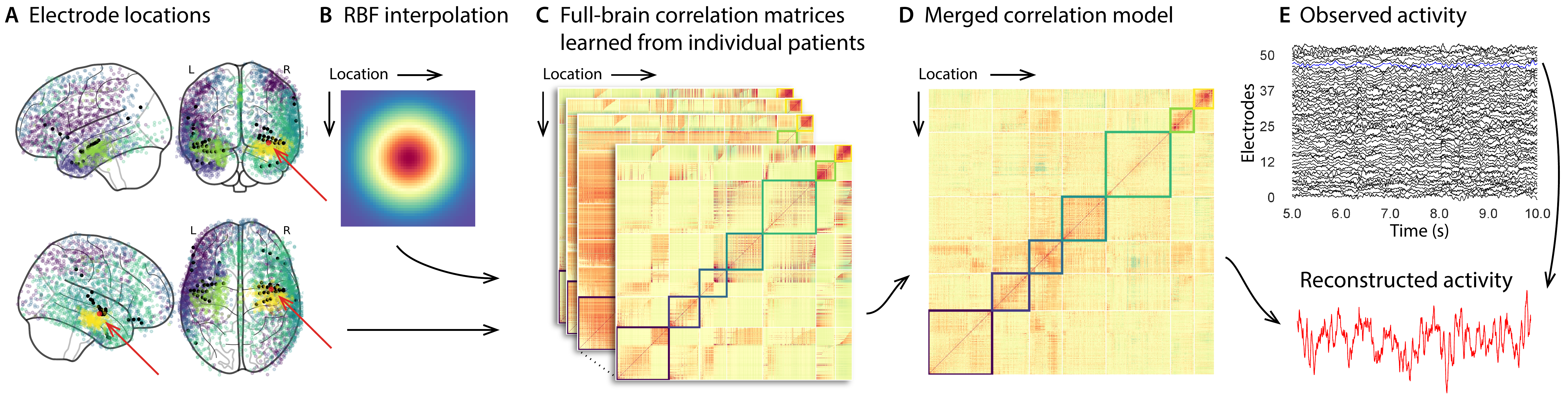}
  \caption{\textbf{Building across-patient models using Gaussian
      process regression.}  \textbf{A.~Electrode locations.}  Each dot
    reflects the location of a single electrode implanted in the brain
    of one patient.  A held-out recording location from one patient is
    indicated in red, and the patient's remaining electrodes are
    indicated in black. The electrodes from the remaining patients are
    colored by $k$-means cluster (computed using the full-brain
    correlation model shown in Panel D).  \textbf{B.~Radial basis
      function kernel.} Each electrode contributed by the patient
    (black) weights on the full set of locations under consideration
    (all dots in Panel A).  The weights fall off with positional
    distance (in MNI152 space) according to a radial basis
    function. \textbf{C.~Per-patient correlation matrices.}  After
    computing the pairwise correlations between the recordings from
    each patient's electrodes, correlations between all locations may
    be estimated using radial basis function-weighted averages.  This
    yields one estimated full-brain correlation matrix for each
    patient.  \textbf{D.~Merged correlation model.}  Combining the
    per-patient correlation matrices (Panel C) yields a single
    full-brain correlation model that captures information contributed
    by every patient.  Here the rows and columns are sorted to reflect
    $k$-means clustering labels~\citep[using $k$=7;][]{YeoEtal11},
    whereby locations are grouped according to their correlations with
    the rest of the brain (i.e., rows of the matrix displayed in the
    panel).  The boundaries denote the cluster groups.  The rows and
    columns of Panel C have been sorted using the Panel D-derived
    cluster labels.  \textbf{E.~Reconstructing activity throughout the
      brain.}  Given the observed recordings from the given patient
    (shown in black; held-out recording is shown in blue), along with
    a full-brain correlation model (Panel D), applying Gaussian
    process regression yields the most probable activity at the
    held-out location (red).  Note: this figure is adapted
    from~\citep{OwenEtal20}, and data are from~\citep{SedeEtal03,
      SedeEtal07a, SedeEtal07b, MannEtal11,
      MannEtal12}.} \label{fig:supereeg}
\end{figure}

\paragraph{Gaussian process models.}  \textbf{Gaussian process
  regression}~\citep{Rasm06} is an approach for estimating ``missing''
(unobserved) data by using related observed data.  Gaussian process
regression is particularly well-suited to applications where nearby
datapoints are expected to take on similar values.  For example, if we
assume that nearby locations in the brain will exhibit similar neural
activity patterns, we could use an approach like Gaussian process
regression to estimate the most probable activity patterns from
locations that were nearby (but not necessarily exactly overlapping
with) the electrode implantation sites for that
patient~\citep{OwenEtal20}.  An overview of this technique is shown in
Figure~\ref{fig:supereeg}.

To build an across-subject model of neural activity patterns, we first
need to define a set of locations in the brain to include in the
combined model (Fig.~\ref{fig:supereeg}A).  These locations comprise
the set of brain coordinates where we will want to estimate activity
patterns for every patient.  Next, we estimate a \textbf{correlation
  model} that describes how activity exhibited by each pair of
locations is related.  To estimate the correlation model, we first
compute the pairwise correlations between activity recorded from each
individual patients' electrodes, and then we use spatial blurring
(Fig.~\ref{fig:supereeg}B) to interpolate those correlations over the
full set of target locations in the model.  This yields a single
estimated correlation matrix for each patient
(Fig.~\ref{fig:supereeg}C).  We can then use weighted averaging to
combine the patient-specific correlation matrices into a single
correlation model (Fig.~\ref{fig:supereeg}D).  Essentially, an entry
in a given patient's individual correlation matrix will be weighted
more heavily in the combined model if that patient had electrodes
nearby to the pair of locations that entry reflects.  Conceptually,
this combined correlation model reflects ``global'' information from
multiple patients, whereas each individual patient's correlation
matrix reflects ``local'' information from that patient alone.  Given
a correlation model (learned from multiple patients) and a set of
recordings (observed from one patient), Gaussian process regression
may be applied to reconstruct (i.e., estimate) neural activity at any
location in the combined model-- even if the given patient did not
have any electrodes at that location (Fig.~\ref{fig:supereeg}E).

Using Gaussian process regression to estimate full-brain activity
patterns from a limited number of electrodes can be useful for
identifying stimulus-driven neural activity.  For example, whereas raw
intracranial recordings are typically taken from different locations
across patients (Fig.~\ref{fig:electrodes}), the above approach may be
used to estimate activity patterns at a common set of locations across
people.  Second, whereas raw intracranial recordings from a single
patient typically lack full-brain coverage
(Fig.~\ref{fig:electrodes}A), the set of locations in the combined
correlation model may be chosen to cover arbitrarily much of the
brain, at arbitrarily high spatial resolution.  In turn, this can
enable researchers to train or apply other across-participant models,
such as pattern classifiers, from different patients' intracranial
recordings~\citep{ScanEtal21}.

\paragraph{Hyperalignment and the shared response model.}
Even when we record across subjects from an (ostensibly) overlapping
set of locations or neural features, individual differences in
stimulus-driven neural responses, behavior, internal representations,
and even neuroanatomy can lead to different observed responses.  When
working with intracranial recordings, where the recording locations
rarely overlap across people, and where non-standard neuroanatomical
traits are relatively common, these factors are even more prevalent.
Hierarchical matrix factorization and Gaussian process models make the
simplifying assumption that different individual's underlying neural
representations are spatially similar.  But what if the same
\textit{functional} representations are reflected by different
\textit{spatial} activity patterns across different people?  Models
that match up neural features primarily according to their spatial attributes
will fail to capture or correctly identify (non-spatial) functional
similarities across individuals.  In contrast, \textbf{functional
  alignment models} attempt to discover functional overlap in neural
activity patterns across individuals, even when the neural features
across those individuals are incompatible or out of spatial alignment.

\textbf{Hyperalignment}~\citep{HaxbEtal11} uses the procrustean
transformation to align the neural trajectories of different
individuals into a common feature space (Fig.~\ref{fig:geometry}).
This entails computing the linear re-combination of neural features
(for each individual) that brings the group's neural trajectories into
the closest point-by-point alignment.  Because the procrustean
transformation is invertible, neural features may be mapped between
different individuals, or between specific individuals and the common
feature space.  The \textbf{Shared Response
  Model}~\citep[SRM;][]{ChenEtal15a} is similar to hyperalignment in
that it provides a means of defining a common neural feature space
that is shared across individuals.  SRM extends hyperalignment by
combining the alignment step with a dimensionality reduction step that
attempts to specifically find a lower-dimensional common neural
feature space.

Although hyperalignment and SRM are most often applied to fMRI
data~\citep{HaxbEtal20}, in principle these models are
modality-independent.  For example, one recent study found that
applying SRM to intracranial recordings, taken taken as patients
watched a movie, revealed a set of shared components that co-varied
with the affective content of the movie~\citep{XieEtal21}.  The study
replicated (using intracranial recordings) several key findings from
related fMRI work~\citep{ChanEtal21a}.  Another recent study, using
intracranial recordings taken from the rodent hippocampus,
demonstrated that hyperaligning neural features across animals enabled
reliable across-individual decoding of the animals' spatial
locations~\citep{ChenEtal21}.

\subsubsection{Inter-subject correlation and inter-subject functional
  correlation.}
\textbf{Inter-subject correlations}~\citep[ISC;][]{HassEtal04} and
\textbf{inter-subject functional correlations}~\citep[ISFC;][]{SimoEtal16}
entail computing the correlations between time-aligned signals
recorded from different individuals as they perform a common task
(Fig.~\ref{fig:patterns}).  ISC operates on the same (or equivalent)
neural features across individuals, and ISFC operates on different
\textit{pairs} of neural features across individuals.

To compute ISC for a particular neural feature, we first isolate that
feature's timeseries in each individual's brain.  (If the same neural
feature is not present across individuals, ISC may be performed after
employing another approach to equating or mapping between neural
features across individuals.)  Next, for a single ``reference''
individual, we correlate their timeseries (for the given neural
feature) with the timeseries for the same neural feature averaged
across all \textit{other} individuals.  This yields a single
correlation coefficient for that individual, for the given neural
feature.  Repeating this calculation using each individual in turn
as the reference yields one correlation coefficient for each
individual.  Finally, we average the correlations across individuals
to obtain a single ISC value for the given neural feature.

Computing ISFC is similar to computing ISC.  However, whereas ISC
correlates the \textit{same} neural feature across individuals, ISFC
correlates \textit{different} neural features across individuals.  The
result is a symmetric matrix of correlation values that summarize how
(on average, across individuals) each pair of neural features are
correlated.

ISC and ISFC are particularly effective at capturing stimulus-driven
activity patterns during naturalistic tasks (e.g., story listening,
movie viewing, natural conversation, etc.) when constructing a
reliable model of the stimulus timecourse can be
challenging~\citep{SimoChan20}.  Effectively, ISC and ISFC treat the
average signals recorded from \textit{other} participants as a
``model'' of the stimulus dynamics.  Since non-stimulus-driven
activity patterns are not expected to be correlated across people, ISC
and ISFC are designed to specifically identify timecourses of
stimulus-driven neural patterns.  Although these approaches are most
commonly applied to non-invasive recordings, they have been
successfully applied to intracranial recordings as
well~\citep{MukaEtal05, HoneEtal12a, PoteEtal14, HaufEtal18}.

\section{Summary and concluding remarks}
Identifying stimulus-driven neural activity requires selecting an
appropriate recording modality and experimental paradigm, defining
neural (Sec.~\ref{sec:activity}) and stimulus
(Sec.~\ref{sec:stimmodel}) features, and then building explicit or
implicit linking functions between neural and stimulus features
(Sec.~\ref{sec:linking}).  We reviewed two general strategies for
building these links: within-participant approaches and
across-participant approaches.

Within-participant approaches include generalized linear models,
multivariate pattern analysis, representational similarity analysis,
and joint stimulus-activity models.  These approaches each attempt to
identify individual-specific maps, patterns, and/or response
profiles.  Across-participant approaches include hierarchical matrix
factorization models, Gaussian process models, geometric alignment
models, inter-subject correlation, and inter-subject functional
correlation.  These approaches each attempt to identify
stimulus-driven neural activity that is similar across individuals.

We also identified several challenges that are unique to intracranial
recordings.  These challenges primarily stem from two factors.  First,
building across-participant models requires accounting for differences
in electrode placement, number of electrodes, and electrode type,
across individuals (Fig.~\ref{fig:electrodes}).  Second, because
intracranial electrodes must be implanted surgically, the subject
population in human intracranial experiments is limited to
neurosurgical patients with serious neurological symptoms such as
drug-resistent epilepsy.  These symptoms often result from brain
abnormalities (e.g., trauma or other forms of physical damage, developmental
abnormalities, and/or other structural or functional issues).  These
issues provide challenges both to comparing findings across
individuals within an intracranial experiment, and also to
generalizing any findings to the broader population.

As a field, cognitive neuroscience is still decades away from being
able to link neural and stimulus features at high levels of detail.
This is partly due to recording quality (even in high-fidelity
modalities like intracranial recordings) and coverage, and partly due
to insufficient quality or fidelity of stimulus models and decoding
algorithms.  Nevertheless, insights into the associations between
stimuli and neural responses can help to elucidate the neural basis of
cognition in ways that behavior alone cannot.  For example, when
behaviors are ambiguous (e.g., a response could convey several
meanings, a response could arise from several equally reasonable or
likely cognitive processes, etc.) or when there \textit{are} no
behaviors for a given cognitive phenomenon (e.g., forgetting, unshared
internal thoughts, etc.), additional signal is needed to resolve those
ambiguities.  In addition, understanding the neural underpinnings of
cognition requires measuring neural activity in some form.
Recent developments in natural language processing and deep learning,
along with advances in tools for more easily constructing neural,
stimulus, and decoding models~\citep[e.g.,][]{GelmEtal15, AbadEtal16,
  PaszEtal19} suggest a bright future for this important area of
neuroscientific inquiry.

\end{document}